\documentclass[apj]{emulateapj}
\usepackage{epsfig}
\usepackage{mathptmx} 
\usepackage{graphicx}

\shorttitle{Galaxy Mergers with Adaptive Mesh Refinement}
\shortauthors{Kim, Wise, \& Abel}

\begin{document}

\title{Galaxy Mergers with Adaptive Mesh Refinement: Star Formation and Hot Gas Outflow}

\author{Ji-hoon Kim \altaffilmark{1,2}}
\email{mornkr@slac.stanford.edu}
\author{John H. Wise \altaffilmark{3}}
\author{Tom Abel \altaffilmark{1,2}}

\altaffiltext{1}{Kavli Institute for Particle Astrophysics and Cosmology, SLAC National Accelerator Laboratory, Menlo Park, CA, USA}
\altaffiltext{2}{Physics Department, Stanford University, Stanford, CA, USA}
\altaffiltext{3}{Laboratory for Astronomy and Cosmology, NASA Goddard Space Flight Center, Greenbelt, MD, USA}

\begin{abstract}

In hierarchical structure formation, merging of galaxies is frequent and known to dramatically affect their properties.   To comprehend these interactions high-resolution simulations are indispensable because of the nonlinear coupling between pc and Mpc scales.  To this end, we present the first adaptive mesh refinement (AMR) simulation of two merging, low mass, initially gas-rich galaxies (${1.8 \times10^{10} M_{\odot}}$ each), including star formation and feedback.   With galaxies resolved by $\sim$$2 \times 10^7$ total computational elements, we achieve unprecedented resolution of the multiphase interstellar medium, finding a widespread starburst in the merging galaxies via shock-induced star formation.  The high dynamic range of AMR also allows us to follow the interplay between the galaxies and their embedding medium depicting how galactic outflows and a hot metal-rich halo form.   These results demonstrate that AMR provides a powerful tool in understanding interacting galaxies.

\end{abstract}


\section{Introduction}

Decades of work have been devoted to the study of interacting and merging galaxies, as they play essential roles not only in shaping present-day galaxies \citep[][``merger hypothesis'']{1977egsp.conf..401T}, but also in constructing large scale structures from the bottom up \citep[][``hierarchical structure formation'']{1978MNRAS.183..341W}.  Because of the nonlinear coupling between pc (star forming regions) and Mpc scales (the distance at which tidal interactions occur) accurate numerical studies are imperative to comprehend the evolution of interacting galaxies.  Although the morphology of merger remnants has been well reproduced by N-body simulations since the pioneering work by \cite{1972ApJ...178..623T}, and various physical characteristics and merger-driven starbursts have been successfully analyzed with smoothed particle hydrodynamics (SPH) simulations \citep{1996ApJ...471..115B, 1996ApJ...464..641M, 2005ApJ...620L..79S, 2005MNRAS.361..776S, 2005Natur.433..604D, 2006ApJ...643..692C, 2006MNRAS.373.1013C, 2007Sci...316.1874M, 2007ApJ...663...61L, 2008arXiv0805.0167S}, a complete, self-consistent simulation of galaxy mergers has not yet been perfected.  

First, SPH simulations tend to have coarse resolution in an interstellar medium (ISM), leading to the over-mixture of different gas phases \citep{2007MNRAS.380..963A, 2008arXiv0808.1844T}.  Therefore, straightforward SPH simulations might have complications in realizing a multiphase medium, in capturing shock-induced star formation, in converting thermal  feedback to a kinetic motion, and thus in showing how feedback makes a difference in galactic evolution self-consistently, though different formulations and sub-resolution models alleviated the problems \citep{2003MNRAS.345..561M, 2003MNRAS.339..289S, 2004MNRAS.350..798B, 2006MNRAS.371.1125S}.

Second, since gaseous halos and an intergalactic medium (IGM) have not been sufficiently resolved in SPH simulations, it is not easy to investigate the interplay between a galactic disk and a diffuse embedding medium.  For instance,  \cite{2006ApJ...643..692C} emphasized that a galactic halo should be included to accurately study the galactic wind and the enrichment process powered by feedback and mergers.  Yet because of their Lagrangian nature and smoothing scheme, SPH simulations might not have sufficient resolution to follow the evolution of the diffuse medium and the galactic outflow.  

In light of these needs, an adaptive mesh refinement (AMR) technique potentially provides a uniquely useful tool to address these issues, allowing us to realize a self-consistent, high-resolution galaxy merger simulation.  As proven by an increasing number of groups \citep{2003ApJ...587...13T, 2008ApJ...672..888T, 2005ApJ...623..650K, 2007arXiv0712.3285C, 2008A&A...477...79D} AMR simulations have been highly successful in resolving the detailed structure of galactic evolution.  In order to make use of the advantages of AMR such as the high dynamic range and reliable shock resolution, we utilize the AMR code {\it enzo} \citep{2001astro.ph.12089B, 2004astro.ph..3044O}.  In this {\it Letter}, we focus on the the first of its kind AMR simulation of two merging, low mass, initially gas-rich galaxies, including star formation and feedback, with emphases on shock-induced star formation and the hot gas outflows.  

\section{Methodology and Numerical Simulations}

\subsection{Data Conversion Pipeline}

We developed a data conversion method which converts a galactic N-body dataset of {\it GalactICS} \citep{1995MNRAS.277.1341K} to an SPH dataset for {\it Gadget} \citep{2001NewA....6...79S}, and then to an adaptive mesh for {\it enzo} employing Delaunay tessellation onto an oct-tree structure.  Using the particle data of a galactic sized halo with both the dark matter and the gas \cite{2008AIPC..990..429K} demonstrated the compatibility of the initial N-body dataset and the adaptive mesh produced through the pipeline.  A suite of functionality checks finds very satisfactory results enabling us to study galaxy evolution with AMR. 

\begin{figure}[t]
\epsscale{1.05}
\plotone{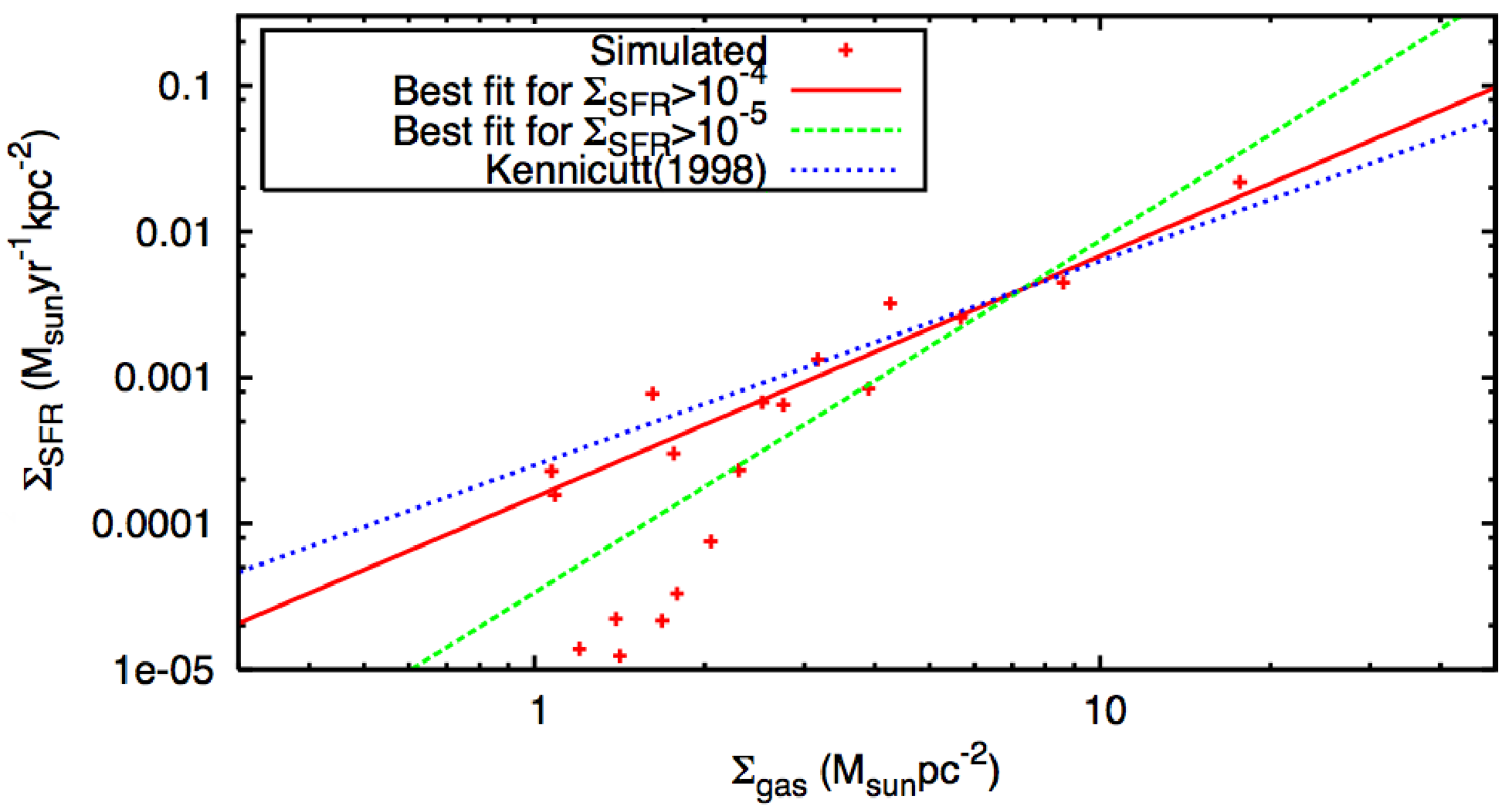}
%\plotone{fig1.png}
\caption{The global K-S relation for an isolated galaxy: time variation of the relationship between global SFR and gas surface density.  Each data point represents a different epoch, equally spaced in 5 Gyr.  The solid line is the best fit for simulated data of $\Sigma_{\rm SFR} > 10^{-4} M_{\odot} {\rm yr^{-1} kpc^{-2}}$, and the dashed line for $\Sigma_{\rm SFR} > 10^{-5} M_{\odot} {\rm yr^{-1} kpc^{-2}}$.  The dotted line is from \cite{1998ApJ...498..541K}.
\label{fig:KS law}}
\end{figure}

\subsection{Simulation Code}

The high-resolution Eulerian AMR code {\it enzo} captures the gravitational collapse of turbulent fragmentation with very high spatial resolution \citep[e.g.][]{2008ApJ...682..745W} and attains multiphase gas dynamics in the ISM as it sharply resolves shocks and phase boundaries \citep{2003ApJ...587...13T, 2005MNRAS.356..737S, 2007MNRAS.380..963A, 2008arXiv0808.1844T}.  {\it Enzo} also contains all relevant physics previously discussed in simulating galaxy evolution processes  \citep{2006ApJ...641..878T, 2008ApJ...673..810T}.

We employ the ZEUS hydrodynamics module included in {\it enzo} to evolve the gas.  Radiative cooling is used by adopting \cite{1987ApJ...320...32S} to follow the equilibrium cooling function down to $10^4$ K,  and \cite{1995ApJ...440..634R} further down to 300 K.  This treatment will ensure a thin galactic disk forms by being cooled below $10^4$ K, the approximate $T_{vir}$ of the ISM in a galactic disk.  The cutoff at 300 K roughly models the temperature floor provided by nonthermal pressure such as cosmic rays and magnetic fields  \citep{1995ApJ...440..634R}.  

Galaxies are placed in a box of 4 Mpc on a side to ensure enough space for galactic tidal interactions and to reduce any boundary effect.  The top grid of $128^3$ cells is allowed to recursively refine up to 13 levels based on the baryonic mass and the dark matter mass in each cell, achieving 3.8 pc resolution in the ISM.  This value is in accord with the Jeans length for a dense gas clump of $n=10^3$ ${\rm cm^{-3}}$, at which a corresponding Jeans mass of $2\times10^3M_{\odot}$ collapses to form a star particle.  In this way, merging galaxies are resolved with $\sim$$2\times10^7$ total computational elements, surpassing any numerical studies conducted thus far on galaxy mergers including gas.

Our star formation criteria are based on \cite{1992ApJ...399L.113C} with several important modifications.  A cell of size $\Delta x$ produces a star particle of $m_{*} = \epsilon \rho_{\rm gas} \Delta x^3$ ($\epsilon$=0.5, a star formation efficiency) when {\it (i)} the gas density exceeds $n_{\rm thres}=10^3$ ${\rm cm^{-3}}$,  {\it (ii)} the flow is converging, {\it (iii)} the cooling time is shorter than the dynamical time, and {\it (iv)} the particle produced has at least $10^3M_{\odot}$.   We do not impose any stochastic star formation unlike \cite{2006ApJ...641..878T} or \cite{2006MNRAS.373.1074S}.  With these revisions, our criteria guarantee that a star particle forms before an unphysically large mass begins to accrete onto any unresolved dense gas clump.  

The energy loss by radiative cooling can be replenished by thermal stellar feedback.  For each star particle, $5 \times 10^{-6}$ of its rest mass energy and 25\% of its mass are returned to the gas over the dynamical time of the particle.  This corresponds to $10^{51}$ ergs per every $110 M_{\odot}$ deposited as stellar mass and represents various types of feedback such as protostellar outflows \citep{2006ApJ...640L.187L}, photoionization \citep{1989ApJ...345..782M}, stellar winds, and Type II supernovae explosions \citep{2006ApJ...641..878T}.  This feedback heats $\sim$$10^3M_{\odot}$ in a $<$10 pc cell up to $\sim$$10^7$ K, but a multiphase medium is naturally established because the cooling time of these hot cells is always much longer than the sound crossing time.  

\begin{figure}[t]
\epsscale{1.2}
\plotone{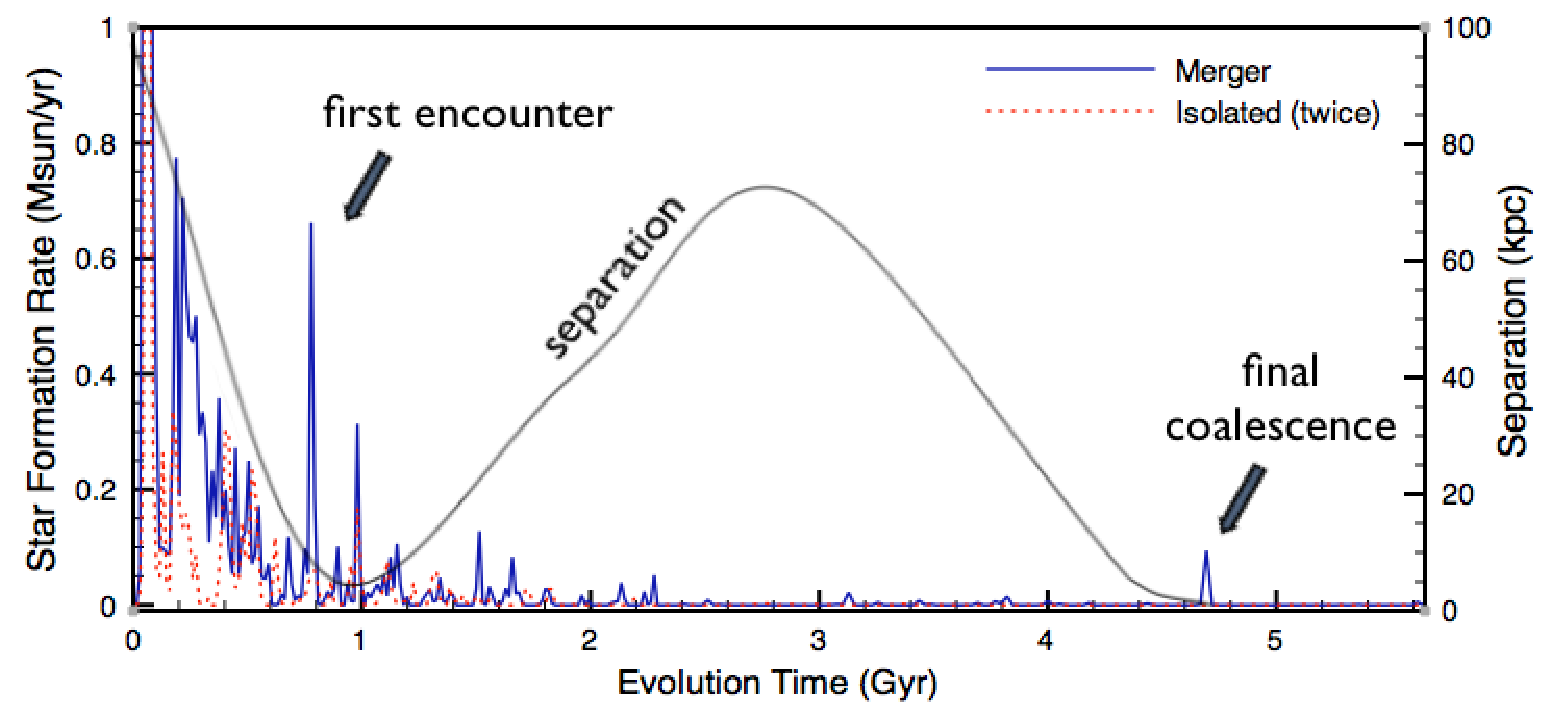}
%\plotone{fig2.png}
\caption{The global SFR as a function of time during the galaxy merger run and the isolated galaxy formation run (twice the value).  The separation between the centers of two galaxies is also displayed.
\label{fig:SFR}}
\end{figure}

\subsection{Initial Conditions}

The individual galaxy progenitor we modeled has a total mass of ${1.8 \times10^{10} M_{\odot}}$ with 10\% in gas ($R_{vir}$ = 65 kpc at $z$=0).  Because we generate gas grids by splitting $10^5$ collisionless particles  in N-body data with the same density profile and velocity dispersion, the gas will virialize to the desired $T_{vir}$ of the galactic halo automatically.  A spin parameter $\lambda$=0.055 is given to cause the progenitor to form a disk galaxy with a gaseous halo within a few hundred Myrs.   In addition, this galaxy progenitor is bathed in a warm ($10^5$ K) diffuse background IGM; an initial metallicity of $10^{-4}Z_{\odot}$ is also set up everywhere to follow the metallicity evolution.  For a merger simulation, two identical galaxy progenitors are separated by 100 kpc and set on a prograde hyperbolic ($e$=1.1) coplanar collision course with a pericentric distance of 4 kpc.  The initial separation is large enough to form individual galaxies before the first passage, and to observe the collision interface between the two gaseous halos.  

\section{Results}

\subsection{Properties of An Isolated Galaxy Model}

We first examine how well our isolated galaxy formation simulation fits the global Kennicutt-Schmidt (K-S) relation between global star formation rate (SFR) and gas surface density, namely $\Sigma_{\rm SFR} \propto \Sigma_{\rm gas}$${^{1.4}}$ \citep{1998ApJ...498..541K}.  To calculate both densities, we select a disk of radius 2.8 kpc so that 95\% of created stars is contained in it after 5 Gyr;  the gas surface density is averaged over the cells where the gas density exceeds $n = 4.0\times 10^{-3} \, {\rm cm^{-3}}$.  Figure \ref{fig:KS law} shows our star formation criteria and feedback correctly match the observed K-S relation, as other simulation works did \citep [e.g.][]{2008ApJ...680.1083R}.  The closest match to the K-S relation occurs when we restrict the fit to $\Sigma_{\rm SFR} > 10^{-4} M_{\odot} {\rm yr^{-1} kpc^{-2}}$, which happens mostly in the first 2 Gyr, following an observational cutoff as in \cite{1998ApJ...498..541K}.

\begin{figure*}[t]
\epsscale{1.1}
\plotone{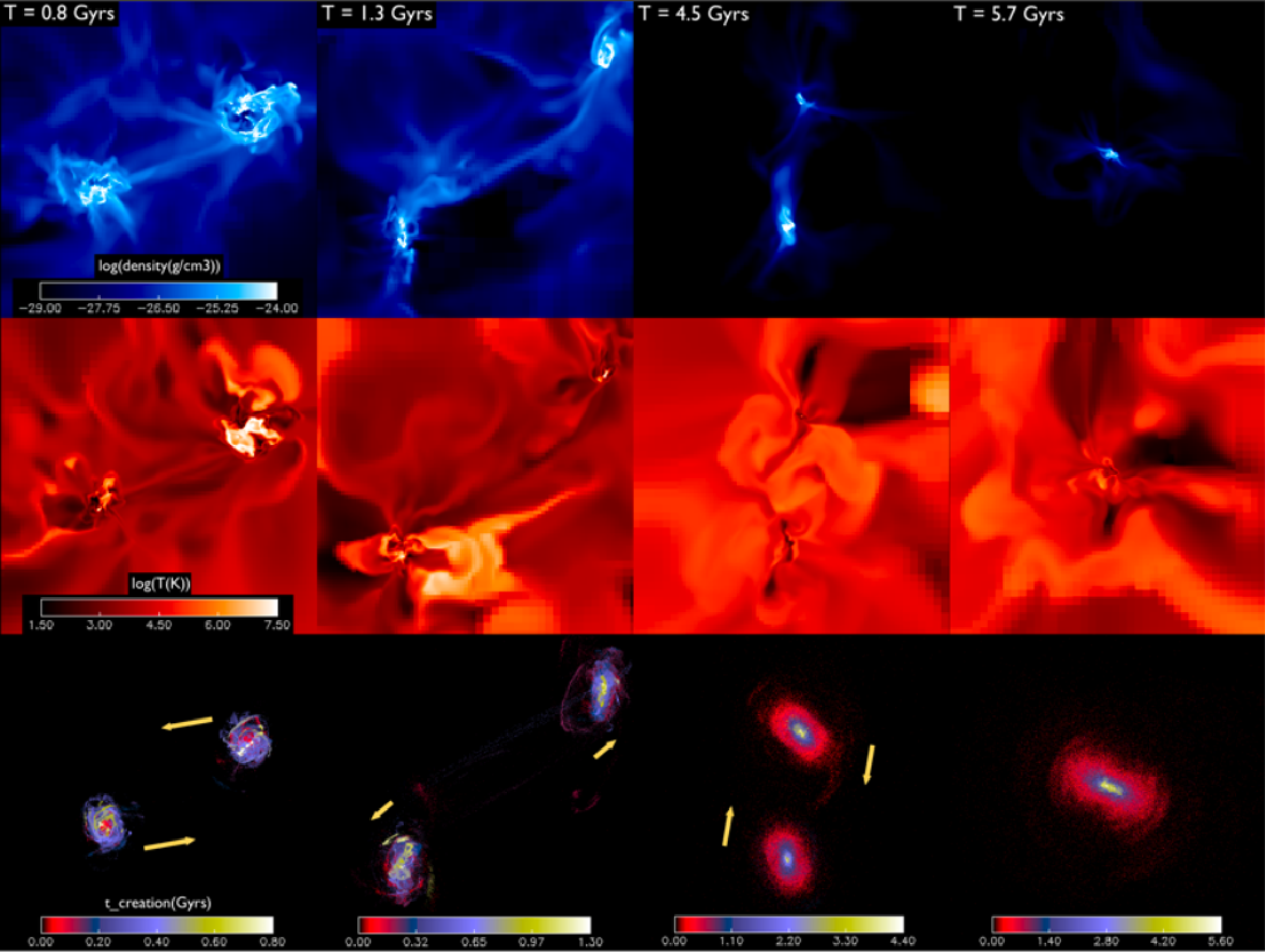}
%\plotone{fig3-1.png}
\caption{Density-weighted projection of density (top), temperature sliced at the orbital plane (middle), and stellar distributions colored by creation time (bottom) in the central 40 kpc, after 0.8, 1.3, 4.5, and 5.7 Gyr.  High-resolution images and movies are available at http://www.slac.stanford.edu/$\sim$mornkr/.
\label{fig:sequence}}
\end{figure*}

\subsection{Star Formation History in A Galaxy Merger}

The global SFR of the merger simulation is displayed in Figure \ref{fig:SFR}.  It presents the initial stellar disk formation for each galaxy in the first $\sim$0.6 Gyr, and several merger-driven starbursts afterwards, notably when two galaxies first encounter ($\sim$0.8 Gyr) and when they finally coalesce to form one galaxy ($\sim$4.7 Gyr).   A low SFR between these two bursts confirms the regulated star formation by stellar feedback.  

Snapshots of the merger sequence at four different epochs are compiled in Figure \ref{fig:sequence}.  The top row shows the density-weighted projection of density, in which irregular gas filaments, bridges and rings are formed by the compression of gas and turbulence.  The middle row depicts the temperature sliced at the collisional plane, where cold gas clumps and hot supernovae bubbles coexist side by side forming a complex, yet well-resolved multiphase medium.  It also reveals how hot supernovae bubbles propagate through the diffuse embedding medium of the halo and the IGM.  In the bottom row of stellar distributions colored by creation time, both merger-induced nuclear starbursts {\it and} shock-induced widespread starbursts are noticeable.  Because of the finer resolution in the ISM, it is easier to resolve local dense clumps driven by shocks and the ensuing star formation.  In contrast, SPH simulations often report predominantly nuclear starbursts \citep{2004MNRAS.350..798B}.  

\subsection {Gas Outflows and Formation of A Hot Gaseous Halo}  

\begin{figure}[t]
\epsscale{1.1}
\plotone{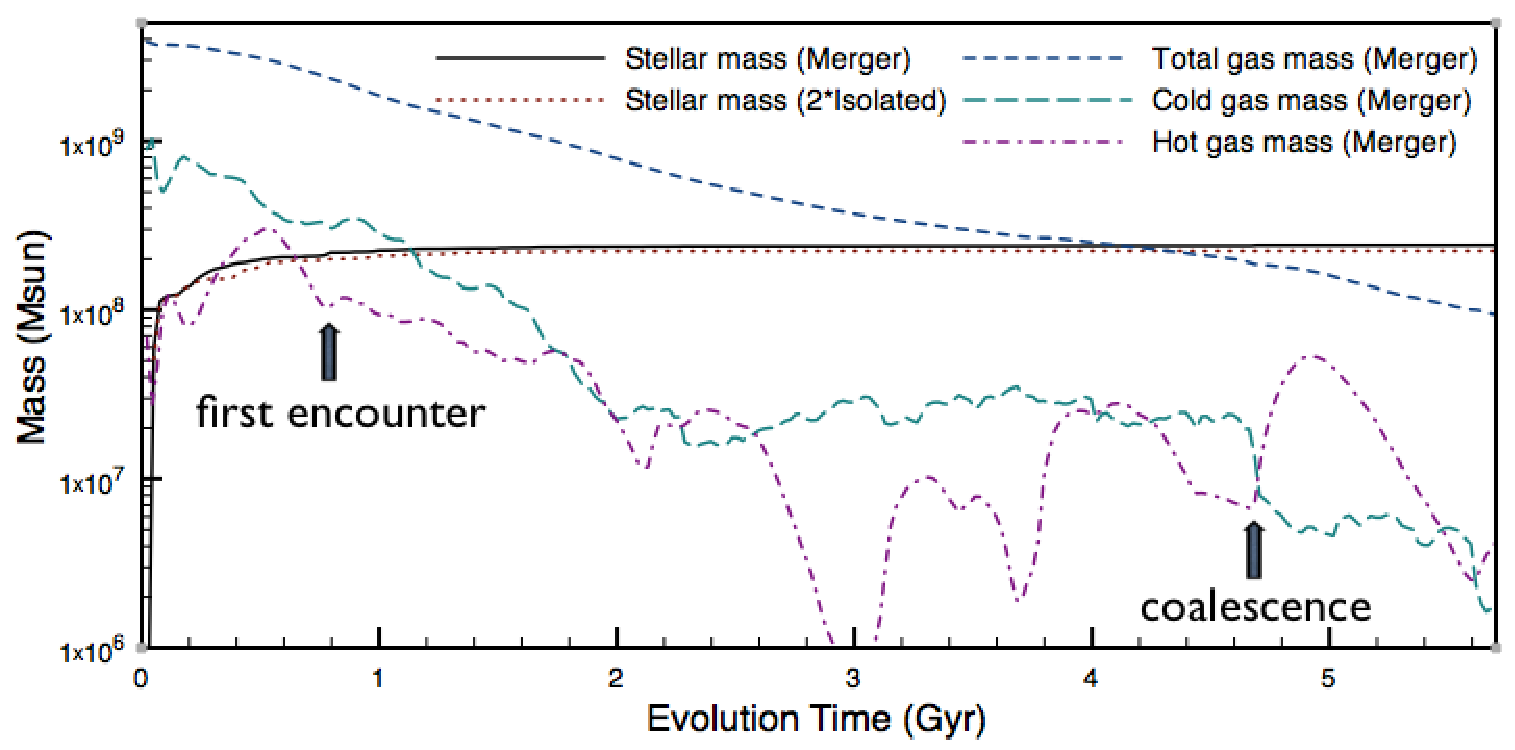}
%\plotone{fig4.png}
\caption{The evolution of the stellar mass (solid line for the merger run and dotted line for twice the value of the isolated galaxy) and the gas mass remaining in the central 200 kpc cube (short dashed line).  Cold (T $< 10^3$ K) and hot (T $> 10^5$ K) gas masses are also shown  (long dashed line and dot-dash line, respectively).    
\label{fig:gas}}
\end{figure}

The evolution of the stellar and gas mass in the central 200 kpc box is plotted in Figure \ref{fig:gas}.   The gas expulsion via stellar feedback and galactic interaction is pronounced as more than 90\% of the gas has been expelled  in the first 4 Gyr.  This gas eventually escapes the gravitational potential of the system or has not had enough time to fall back onto the galaxies.  This massive gas depletion is prominent especially in low mass mergers because of the shallow gravitational well.  As for the merger remnant at 5.7 Gyr, the remaining gas mass is $\sim$35\% of the stellar mass and still decreasing rapidly.  The amount of cold gas (T $< 10^3$ K) available for future star formation is only $<$1\% of the stellar mass, depicting how star formation is quenched by feedback heating and gas expulsion.  

\begin{figure}[t]
\epsscale{1.22}
\plotone{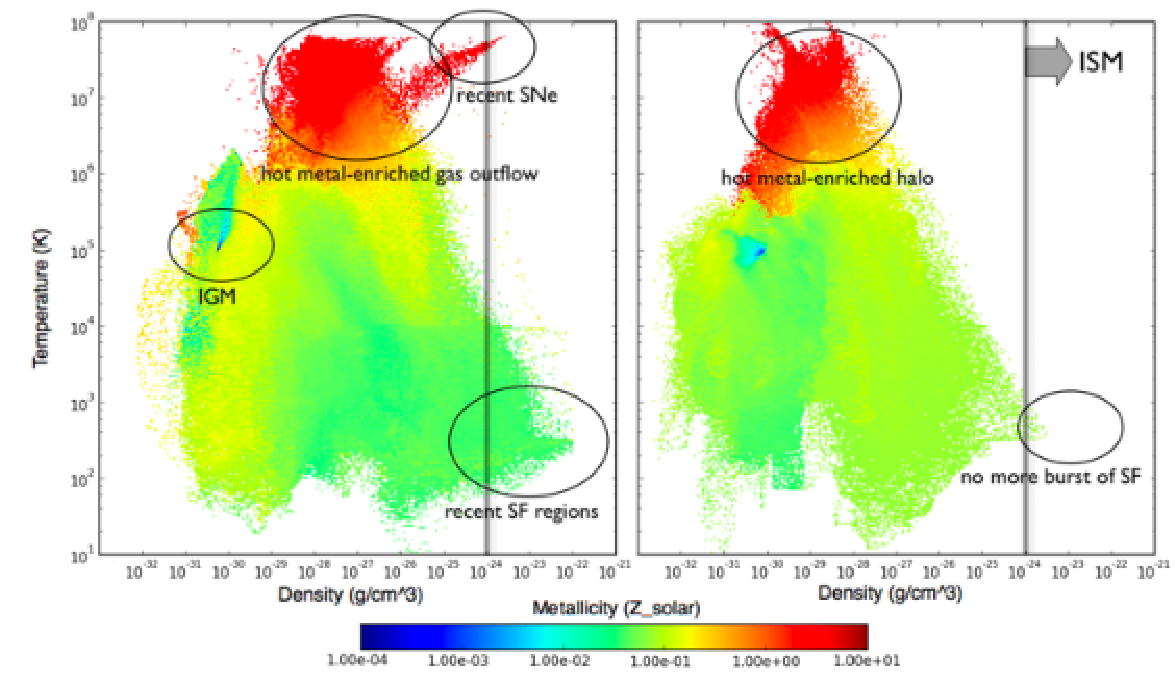}
%\plotone{fig5.png}
\caption{Joint density-temperature PDFs, colored by metallicity, for a 400 kpc sphere centered on the galaxy, after 0.8 Gyr (left) and 5.7 Gyr (right).  Star forming regions, supernovae bubbles, gas outflows, and the halo are pointed out, proving the wide range of densities and temperatures followed here.
\label{fig:PDF}}
\end{figure}

The gas disrupted by galactic interaction and heated by feedback creates a galactic wind of $>$200 km/s reaching as far as 1 Mpc from the simulated merging galaxies.  This hot metal-loaded outflow is responsible for building the gaseous halo around galaxies as well as enriching some regions of the IGM up to a supersolar metallicity.  As a result, a hot metal-rich halo is generated ($\rho \sim10^{-29} {\rm g cm^{-3}}$, T $\sim10^{6-7}$ K) and sustained by continuous stellar feedback, as suggested by analytic models \citep[e.g.][]{2008arXiv0803.4215T}.  Although the galactic outflows and the halo are very diffuse, their evolution is easily followed in AMR, as can be clearly seen in Figure \ref{fig:PDF} of the joint probability distribution functions (PDFs) on density-temperature planes.  It also illustrates the wide range of densities and temperatures that are followed here.

\section{Discussion and Summary}

Our simulation, for the first time, followed the self-consistent evolution of low mass merging galaxies with AMR at unprecedented resolution.  Our findings are as follows.

First, as AMR naturally establishes a multiphase medium without any sub-resolution model, we have captured shock-induced star formation that occurs when merging galaxies compress the intervening gas \citep{2004MNRAS.350..798B, 2008arXiv0805.0167S}.  The well-resolved shocks trigger a widespread starburst, in accord with observations \citep[e.g.][]{2006astro.ph..6036S}.  Further, the overcooling problem is absent as in \cite{ 2007arXiv0712.3285C}  because the multiphase medium is resolved by $<$10 pc cells, and the thermal feedback is sufficient to heat such small cells up to $\sim$$10^7$ K.

Second, utilizing the high dynamic range and the Eulerian nature of AMR, we have followed the evolution of the hot diffuse medium of gaseous halos and the IGM as far as 1 Mpc away from the galaxies.  This allows us to explore the interplay between the galactic outflows and the embedding medium and to demonstrate that a hot metal-rich halo forms around the galaxies from stellar feedback  \citep{2006ApJ...643..692C}. The massive gas expulsion in low mass merging galaxies leads to a high mass-to-light ratio, as it creates a merger remnant without much cold gas left for later star formation.

Although it should be considered provisional, our result brings compelling evidence that AMR delivers a uniquely powerful tool in understanding merging galaxies, while it addresses several issues SPH has suffered from.  Comprehensive parameter studies should follow, especially in the efficiency of stellar feedback and the metal yields of stars; the results should be compared and calibrated with observations such as the mass-metallicity relation \citep{2004ApJ...613..898T}, galactic outflows \citep{2006ApJ...647..222M}, galactic morphology \citep{2008arXiv0809.2156P}, and gas to stellar mass ratio \citep{2008A&A...482...43G}.  Physics such as UV photoelectric heating, cosmic rays, and magnetic fields are missing in this work, but will need to be considered in the future.

\acknowledgments

We thank Marcelo Alvarez, T. J. Cox, Andres Escala, Kristian Finlator, Yuexing Li, Changbom Park, Cecilia Scannapieco, Volker Springel, Elizabeth Tasker, Matthew Turk, Peng Wang, and Risa Wechsler for insightful comments and advice.   JK was supported by William R. and Sara Hart Kimball Stanford Graduate Fellowship.  JHW was supported by an appointment to the NASA Postdoctoral Program, administered by Oak Ridge Associated Universities.  This work was partially supported by NASA ATFP grant NNX08AH26G and NSF AST-0807312.   These calculations were performed on 32 processors of the Orange cluster at KIPAC.

\end{document}